\documentclass[%
reprint,
 amsmath,amssymb,
 aps,
]{revtex4-1}

\usepackage{graphicx}
\usepackage{dcolumn}
\usepackage{bm}

\begin{document}

\preprint{APS/123-QED}

\title{
Thermal Hall conductivity in the spin-triplet superconductor with broken time-reversal symmetry }%

\author{Yoshiki Imai$^{1}$}%
\email{imai@phy.saitama-u.ac.jp}
\author{Katsunori Wakabayashi$^{2}$}%
\author{Manfred Sigrist$^{3}$}
\affiliation{$^{1}$Department of Physics, Saitama University, Saitama 338-8570, Japan}
\affiliation{$^{2}$Kwansei Gakuin University, Sanda 669-1337, Japan}
\affiliation{$^{3}$Institut f\"ur Theoretische Physik, ETH-Z\"urich, CH-8093 Z\"urich, Switzerland}

\date{\today}

\begin{abstract}
Motivated by the spin-triplet superconductor Sr$_2$RuO$_4$, the thermal Hall conductivity is investigated for several pairing symmetries with broken time-reversal symmetry. 
In the chiral $p$-wave phase with a fully opened quasiparticle excitation gap,  the temperature dependence of the thermal Hall conductivity has a temperature linear term associated with the topological property directly, and an exponential term, which shows a drastic change around the Lifshitz transition. 
Examining $f$-wave states as alternative candidates with $\bm d=\Delta_0\hat{z}(k_x^2-k_y^2)(k_x\pm ik_y)$ and $\bm d=\Delta_0\hat{z}k_xk_y(k_x\pm ik_y)$ with gapless quasiparticle excitations, 
we study the temperature dependence of the thermal Hall conductivity, where for the former state the thermal Hall conductivity has a quadratic dependence on temperature, originating from the linear dispersions, in addition to linear and exponential behavior. The obtained result may enable us to distinguish between the chiral $p$-wave and $f$-wave states in Sr$_2$RuO$_4$. 
\begin{description}
\item[PACS numbers]
\end{description}
\end{abstract}

\pacs{Valid PACS appear here}
\maketitle

\section{Introduction}
The transition metal oxide Sr$_2$RuO$_4$~\cite{maeno94,mackenzie03,maeno12} attracts much attention 
as a candidate of the topological superconductors. 
The superconducting state has spin-triplet Cooper pairing~\cite{ishida98} with broken time-reversal symmetry~\cite{luke98, xia06}. 
Experimental studies suggest that the leading candidate of the superconducting order parameter is the so-called chiral $p$-wave Cooper pairing~\cite{rice95} written as 
\begin{align}
\bm d=\Delta_0\hat{z}(k_x+i\eta k_y), 
\end{align}
where $\eta=\pm 1$ represents the chirality corresponding to the sign of an orbital angular momentum $L_z = \pm 1$ along the $z$ axis. 
This is a two-dimensional analog of the Anderson-Brinkman-Morel (ABM) state of superfluid $^3$He. 
Since Sr$_2$RuO$_4$ has a strong two-dimensional structure, the chiral $p$-wave state has a nodeless quasiparticle excitation gap, and the topological property is characterized by the so-called Chern number. 

On the other hand, the following order parameters with the $f_{x^2-y^2}$-wave and the $f_{xy}$-wave symmetries~\cite{hasegawa00} have also been proposed, and the order parameters are given by
\begin{align}
\bm d=\left\{
\begin{array}{cc}
\Delta_0\hat{z}(k_x^2-k_y^2)(k_x+i\eta k_y) &(f_{x^2-y^2})\\
\Delta_0\hat{z}k_xk_y(k_x+i\eta k_y)&(f_{xy})
\end{array}
\right.,
\end{align}
where there exist line nodes along $ [1,\pm1] $ for the $f_{x^2-y^2}$-wave state, 
and $ [1,0] $ and $ [0,1] $ for the $f_{xy}$-wave state. 
These $f$-wave pairing states lead to the gapless excitations of quasiparticles. 

Angle-resolved photoemission spectroscopy (ARPES), de Haas-van Alphen measurement studies~\cite{damascelli00,mackenzie96,bergemann00,veenstra13} and first principles calculations~\cite{mazin97,imai14} indicate that the low-energy electronic structure of Sr$_2$RuO$_4$ is well described by the Ru 4$d$ $t_{2g}$ orbitals. The Fermi surface consists of the one-dimensional $\alpha$ and $\beta$ bands, and the two-dimensional $\gamma$ band. The latter derives from the $d_{xy}$ orbital mainly,  with the electronlike Fermi surface slightly below the van Hove points [$\bm k=(\pi/a,0)$ and $(0,\pi/a)$] with a lattice constant $a$ which we will set equal to 1 in the following.
Thus the Fermi surface structure is affected by the physical and/or chemical tunings, such as surface distortion, La doping and strain effects~\cite{matzdorf00,kikugawa04A, kikugawa04B,kittaka10,hicks14, taniguchi15, burganov16,liu16, steppke17}. 
These modifications may lead to a Lifshitz transition in the $\gamma$ band near the van Hove points, which is closely connected with the topological property characterized by the Chern number~\cite{imai13}. 

In the chiral $p$-wave phase, while the chiral edge current expected theoretically~\cite{matsumoto99,furusaki01} is nearly insensitive to the presence of the  Lifshitz transition~\cite{imai14}, the thermal Hall effect is more suitable probe to study the topological aspect of the superconducting phase. 
The thermal Hall conductivity is directly proportional to the Chern number of the topological superconductors in the very low-temperature region~\cite{sumiyoshi13,scaffidi15}. 
Assuming the chiral $p$-wave phase, we have investigated the temperature dependence of the thermal Hall conductivity by means of the single band tight-binding model describing the $\gamma$ band, which consists of the temperature linear term and the exponential term~\cite{imai16}. The coefficients of the exponential term lead to information on the Chern number and the excitation gap. 

The low-energy properties including the topological aspect in the $f$-wave phases are different from that in the chiral $p$-wave phase, which is strongly reflected in the thermal Hall conductivity. 
Thus in this paper, we will investigate the temperature dependence of the thermal Hall conductivity for the $f_{x^2-y^2}$-wave and the $f_{xy}$-wave states in addition to the chiral $p$-wave state, which may give us information concerning the distinction between the chiral $p$-wave and the other $f$-wave phases.


\section{Model}

The holelike $\alpha $ and the electronlike $\beta $ bands have nearly one-dimensional character and their net Chern numbers cancel to zero making them insensitive to the Lifshitz transition~\cite{imai12,imai13}.  
On the other hand, the topological properties associated with two-dimensional electronlike $\gamma$ band are strongly affected when the chemical potential traverses the van Hove level, which leads to a qualitative change of some physical quantities. 

Since the thermal Hall conductivity is closely related to the topological aspect in the spin-triplet superconducting phase with broken time-reversal symmetry, we focus on the $\gamma$ band for simplicity. 
By means of the single band tight-binding model on the two-dimensional square lattice, the following Bogoliubov de-Gennes (BdG) Hamiltonian is introduced as 
\begin{align}
H=\sum_{\bm k}
\left(
c^\dag_{\bm k\uparrow }, c_{-\bm k\downarrow }
\right)
\left(
\begin{array}{cc}
\varepsilon_{\bm k}&\Delta_{\bm k}\\
 \Delta_{\bm k}^*&-\varepsilon_{\bm k}
\end{array}
\right)
\left(
\begin{array}{c}
c_{\bm k\uparrow }\\
c^\dag_{-\bm k\downarrow }
\end{array}
\right), 
\label{ham}
\end{align}
where $c_{\bm k\sigma}$ is the annihilation operator of an electron with wave vector $\bm k[= (k_x,k_y )]$ and spin $\sigma(=\uparrow,\downarrow )$. 
$\varepsilon_{\bm k}$ and $\Delta_{\bm k}$ are the energy dispersion of electrons in the normal phase and the gap function, respectively, written as 
\begin{align}
\varepsilon_{\bm k}=-2t(\cos k_x+\cos k_y)-4t'\cos k_x \cos k_y-\mu,
\end{align}
and
\begin{align}
\Delta_{\bm k}=\left\{
\begin{array}{lc}
\Delta (\sin k_x+i\eta\sin k_y),& \text{(chiral $p$)}\\
\Delta (\cos k_x-\cos k_y)(\sin k_x+i\eta\sin k_y), &\text{($f_{x^2-y^2}$)}\\
\Delta \sin k_x \sin k_y(\sin k_x+i\eta\sin k_y), &\text{($f_{xy}$)}
\end{array}
\right.
\label{eqn:defgap}
\end{align}
where $t$ ($t'$) stands for the hopping amplitude between nearest-neighbor (next-nearest-neighbor) sites. 
$\mu$ denotes the chemical potential and $\Delta$ is a superconducting gap amplitude.  

\begin{figure}[tb]
\begin{center}
\includegraphics[height=40mm]{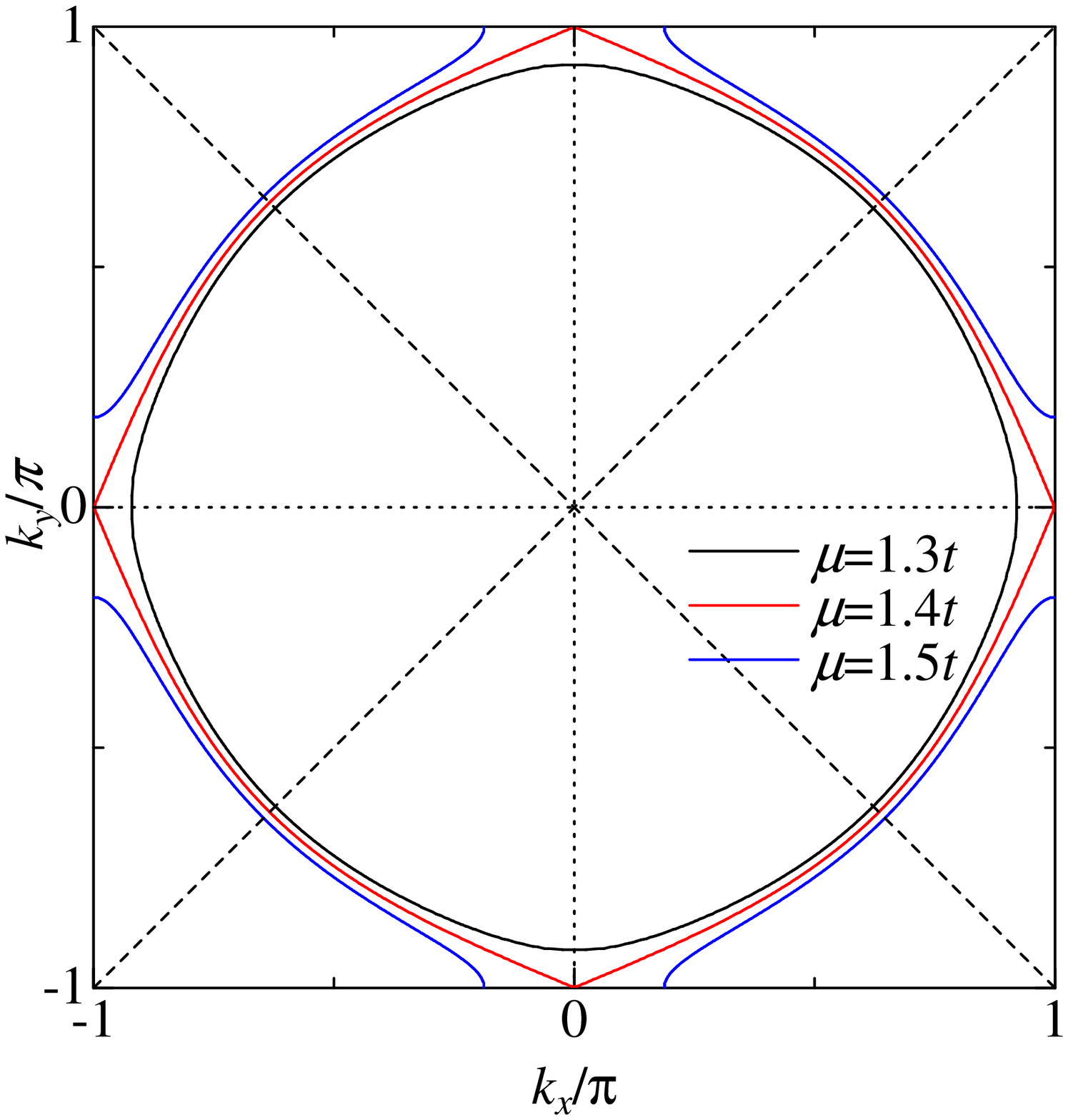}
\caption{(Color online) Fermi surface in the normal state for several choices of $\mu$ with $t'=0.35t$. The dashed (dotted) line stands for the line nodes at $k_x=\pm k_y$ ($k_x=0,\pi$ and $k_y=0,\pi$).  }
\label{FS}
\end{center}
\end{figure}

Figure \ref{FS} displays the Fermi surface of the $\gamma$ band in the normal phase for several choices of the chemical potential $\mu$. 
With increasing chemical potential the Fermi surface topology changes from the electronlike to the holelike one. 
The critical chemical potential is $\mu=1.4t\,(\equiv\mu_c)$. 
The chemical potential will be taken as our parameter to change the Fermi surface topology.

Diagonalizing Eq. (\ref{ham}), we obtain the quasiparticle energy 
\begin{align}
E_{\bm k\pm}=\pm\sqrt{\varepsilon_{\bm k}^2+|\Delta_{\bm k}|^2},  
\end{align}
where the line nodes in the $f$-wave phases generate the gapless excitations. 

At the van Hove points $\bm k=(\pi,0)$ and $\bm k=(0, \pi)$, the gap functions vanish for all pairings in Eq. (\ref{eqn:defgap}) with the quasiparticle energy $E_{\bm k}=4t'-\mu$. 

\section{Results}
We now discuss the temperature dependence of the thermal Hall conductivity, using the model parameters $t'=0.35t$ and $\Delta=0.1t$, and assume the chirality with $\eta=+1 $. 

\subsection{Quasiparticle energy dispersions}
We first study the density of states (DOS) and the quasiparticle energy dispersions around the gapless points for the $f$-wave phases, where 
the excitation gaps vanish at $\bm k=\bm k_0[\equiv (k_0,k_0)]$ in the $f_{x^2-y^2}$-wave phase, and at $\bm k=\bm k_1[\equiv (k_1,0)]$ ($\mu<\mu_c$) or $\bm k=\bm k_2[\equiv (\pi,k_2)]$ ($\mu>\mu_c$) in the $f_{xy}$-wave phase, which are given by
\begin{align}
\begin{array}{lc}
k_0=\arccos \left( \frac{-t+\sqrt{t^2-t'\mu}}{2t'}\right),& (f_{x^2-y^2}\text{-wave})\\
k_1=\arccos \left(-\frac{2t+\mu}{2t+4t'}\right),& (f_{xy}\text{-wave}, \mu<\mu_c)\\
k_2=\arccos \left(-\frac{2t-\mu}{2t-4t'}\right),& (f_{xy}\text{-wave}, \mu>\mu_c). 
\end{array}
\end{align}

\begin{figure}[bt]
\begin{center}
\includegraphics[height=30mm]{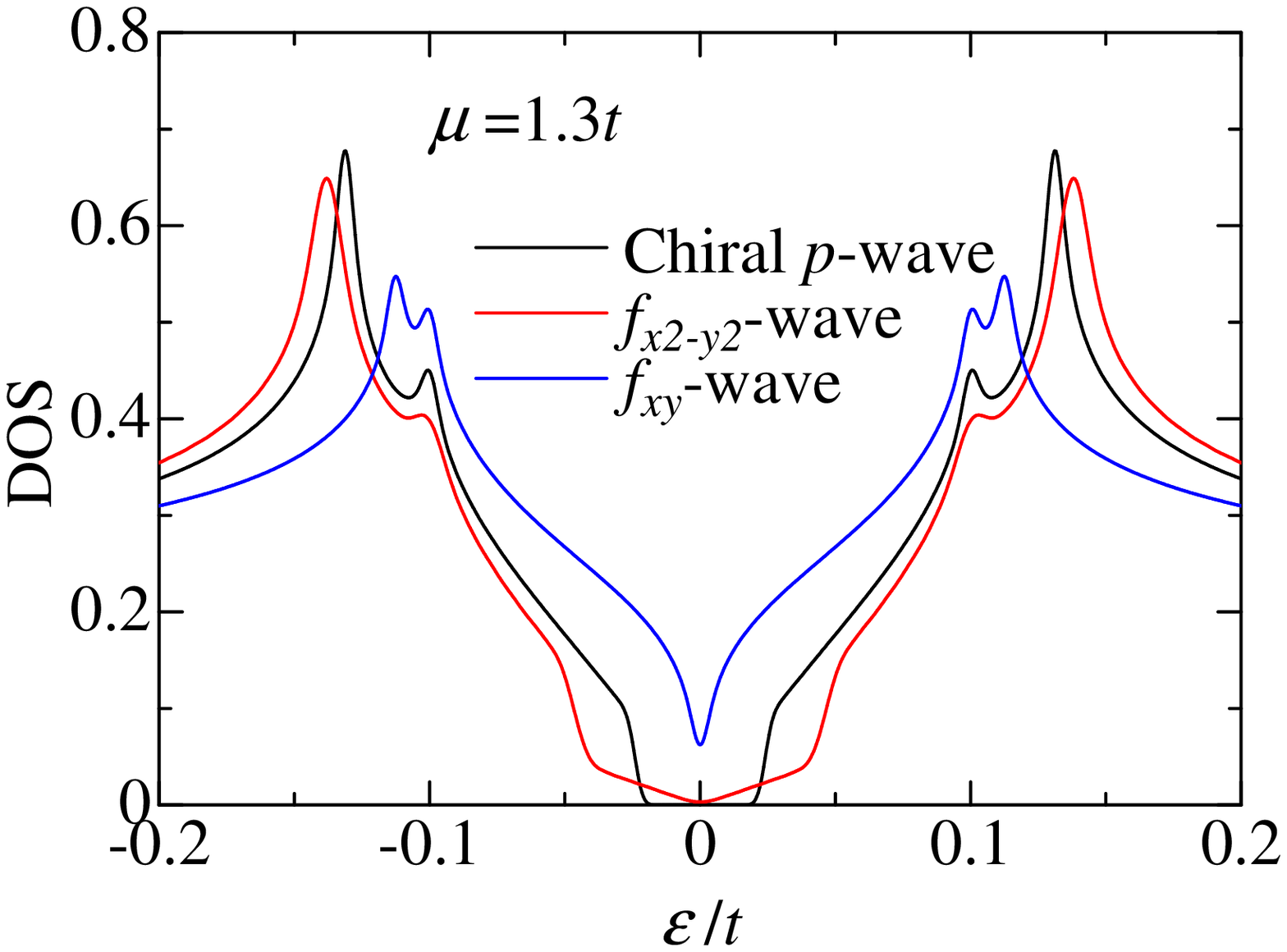}
\includegraphics[height=30mm]{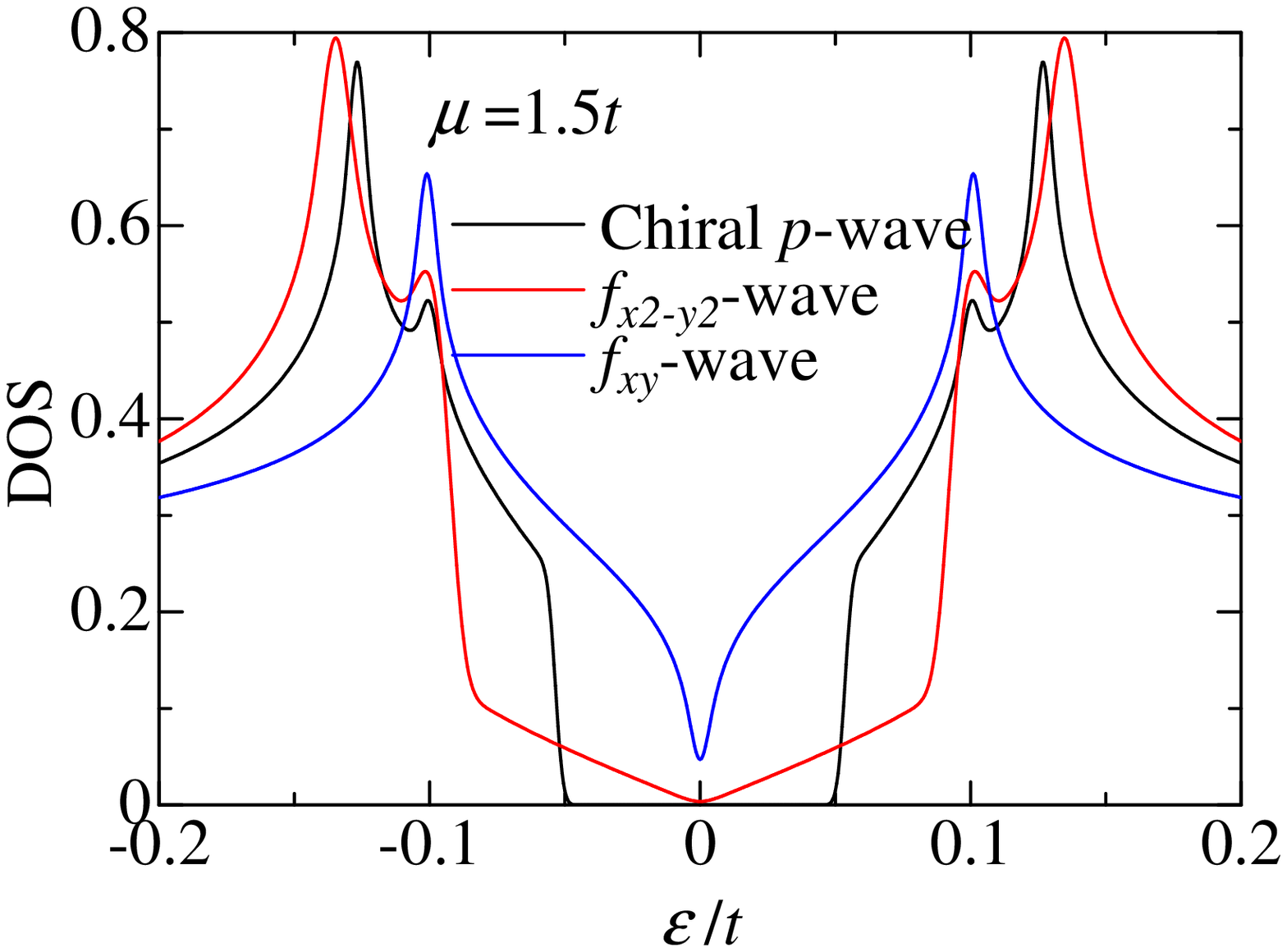}
\caption{(Color online) The bulk quasiparticle DOS at $\varepsilon \sim0$ with $\mu=1.3t$ (left panel) and $\mu=1.5t$ (right panel) for the chiral $p$-wave, the $f_{x^2-y^2}$-wave, and the $f_{xy}$-wave states, respectively.  }
\label{DOS}
\end{center}
\end{figure}

Figure \ref{DOS} displays the bulk DOS in these superconducting phases. 
In the chiral $p$-wave phase there exists a full excitation gap, except for $\mu=\mu_c$.  
Although the $f_{x^2-y^2}$-wave state has a similar DOS structure, it has a pseudogap with excitations to arbitrarily low energy. 
For the $f_{xy}$-phase we find a small finite DOS around $\varepsilon = 0$.
These finite DOSs in the $f$-wave phases are attributed to the structure of the low-energy quasiparticle dispersion, which are expanded in a power series of wave vector around the gapless points. 

In the $f_{x^2-y^2}$-wave phase the lowest order with respect to the $\bm q\equiv \bm k-\bm k_0$ in the energy dispersions reads 
\begin{align}
%
E_{\bm k+}&=\sqrt{a+b\sin \theta}\,q,\\
a&\equiv 2\sin^2 k_0 \{2(t+2t'\cos k_0)^2+\Delta^2\sin^2k_0\},\\
b&\equiv 2\sin^2 k_0 \{2(t+2t'\cos k_0)^2-\Delta^2\sin^2k_0\},
\end{align}
where $\theta$ represents an angular dependence of wave vector $\bm q$ and $q\equiv |\bm k-\bm k_0|$.  
Thus, $E_{\bm k \pm}$ has a linear dispersion for all directions around $\bm k=\bm k_0$, which leads to a linear DOS at $\varepsilon \sim 0$, as shown in Fig. \ref{DOS}. 

On the other hand, the $f_{xy}$-wave state has a more complicated dispersion relation. 
In particular, along the $k_x$ or $k_y$ direction, the energy dispersions for $\mu<\mu_c$ are given by
\begin{align}
E_{\bm k+}&= \left\{
\begin{array}{cc}
c |q_x|, &  \text{along the $k_x$ direction at $k_y=0$} \\
d q_y^2, & \text{along the $k_y$ direction at $k_x=k_1$} 
\end{array}
\right., \\
c&\equiv 2(t+2t')\sin k_1,\>d\equiv (t+2t'\cos k_1), 
\label{eqn:ek_fxy}
\end{align}
with $\bm q=(q_x,q_y)\equiv \bm k-\bm k_1$.
For $\mu>\mu_c$ case, a similar behavior is obtained.  
Because of the contribution from the $q_y^2$ term the finite DOS appears around $\varepsilon =0$ (Fig. \ref{DOS}). 

\begin{figure}[bt]
\begin{center}
\includegraphics[height=101mm]{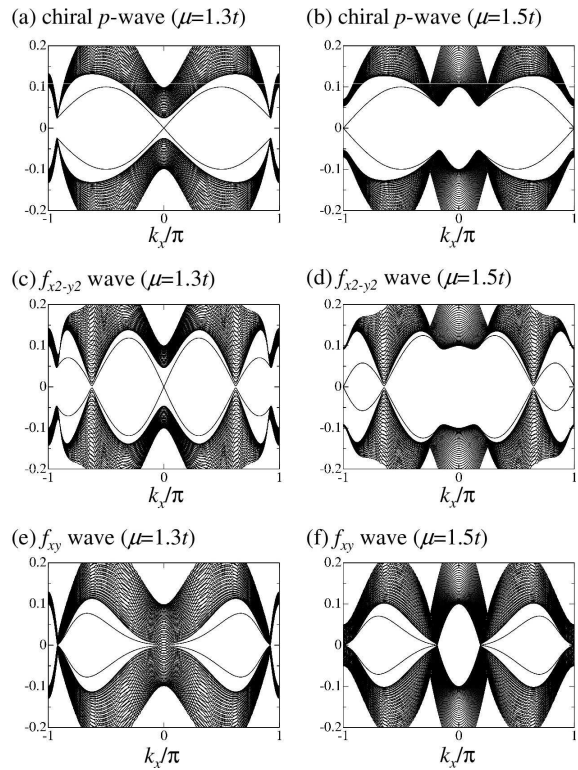}
\caption{Energy dispersions with open boundary conditions in the $y$ direction with $\mu=1.3t$ (left panels) and $\mu=1.5t$ (right panels) for the chiral $p$-wave, the $f_{x^2-y^2}$-wave, and the $f_{xy}$-wave states, respectively.  The number of legs is $L_y=100$. }
\label{band_ribbon}
\end{center}
\end{figure}

We consider now the energy dispersion of a ribbon with open boundary conditions in the $y$-direction, while keeping translation invariance along $x$-direction, which leads to edge states related to the bulk topological properties due to the bulk-edge correspondence~\cite{wen92,hatsugai93} (see  Figure \ref{band_ribbon} )
The number of legs is $L_y=100$, which is sufficiently large to obtain independent edge states at the two edges, while bulk properties are reproduced around the center of the ribbon. 
For all pairing symmetries we find edge states in the low-energy region. 

Linear dispersions for edge states appear in the chiral $p$-wave phase at $k=0$ ($\mu=1.3t$) and $k=\pi$ ($\mu=1.5t$), respectively, whereby the crossing point at $\varepsilon =0$ jumps when passing through the Lifshitz transition. Due to the full excitation gap, these edge states are topologically protected and robust against temperature effect and disorders. 

While there exist similar edge states for the $f_{x^2-y^2}$-wave phase, the bulk superconducting excitation gaps vanish at $k=k_0\sim \pm 0.6\pi$. 
For the $f_{xy}$-wave phase, edge states with crossing $\varepsilon =0$ appear at $k=0,k_1$ ($\mu=1.3t$) and $k=k_2,\pi$ ($\mu=1.5t$), where the bulk gap vanishes too.

\subsection{Chern number}

The non-trivial topology of the superconducting states is captured through the Chern number. The formula is given by~\cite{thouless82,kohmoto85,volovik85}
\begin{align}
N_C&=\frac{4\pi}{M}\sum_{\bm k (E_{\bm kn}<0)}\sum_{n=\pm}{\rm Im}\left\langle \frac{\partial u_{\bm kn}}{\partial k_x}\Bigg|\frac{\partial u_{\bm kn}}{\partial k_y}\right\rangle, \\
\label{eqn:Nc}
&=-2\pi\frac{i}{M}\sum_{\bm k}\sum_{n\ne n'}
\langle n |J_{\bm k}^{x}|n' \rangle\langle n' |J_{\bm k}^{y}|n \rangle
\frac{f(E_{\bm kn})-f(E_{\bm kn'})}{(E_{\bm kn}-E_{\bm kn'})^2},
\end{align}
where $M$ stands for the number of sites and $\bm u_{\bm kn}$ is the periodic part of the Bloch wave function of the BdG equations. 
The matrix element $J_{\bm k}^\mu$ is formally defined like a current ($=\partial H_{\bm k}/\partial k_{\mu}$), and $f (E_{\bm kn})$ is the Fermi-distribution function.  
For the materials with a nodeless insulating or superconducting gap, the Chern number is an integer.  
In order to avoid a divergence,  a very small imaginary constant $i\xi $ is introduced in the denominator of the integrand as follows, 
$(E_{\bm kn}-E_{\bm kn'})^2\rightarrow \lim_{\xi\rightarrow 0}\text{Re}\{(E_{\bm kn}-E_{\bm kn'})^2+i\xi \}$. 

\begin{figure}[bt]
\begin{center}
\includegraphics[height=40mm]{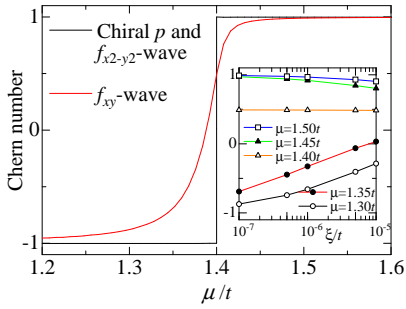}
\caption{(Color online) Chern number as a function of the chemical potential $\mu$ for the chiral $p$-wave ($f_{x^2-y^2}$-wave) and the $f_{xy}$-wave states at absolute zero temperature with $\xi=1.0\times 10^{-7}t$. The inset displays the $\xi$ dependence of $N_C$ for the $f_{xy}$-wave state. }
\label{NT}
\end{center}
\end{figure}
Figure \ref{NT} displays the Chern number as a function of the chemical potential at absolute zero temperature. 
$N_C$ shows the discontinuous change at $\mu=\mu_c$ for the chiral $p$-wave and the $f_{x^2-y^2}$-wave phases, which indicates that the topological property changes concomitantly with the Lifshitz transition. 

The magnitude of the Chern number with integer corresponds to the number of linearly dispersing edge states due to the bulk-edge correspondence, which is consistent with the energy dispersions in Fig. \ref{band_ribbon}. 
Note that while $N_C$ in the chiral $p$-wave phase is insensitive to temperature, $N_C$ in the $f_{x^2-y^2}$-wave phase deviates gradually from integer with increasing temperature. 

The Chern number in the $f_{xy}$-wave phase is sensitive to the magnitude of $\xi $ and moves toward integer upon decreasing $\xi$ except for $\mu\sim \mu_c$. However, in contrast to the chiral $p$-wave and the $f_{x^2-y^2}$-wave cases, the Chern number deviates from integer over a wider range and changes smoothly around $\mu=\mu_c$ with $\xi \ne 0$, even at vanishing temperature, which indicates that edge states are fragile due to the finite DOS at $\varepsilon \sim 0$ resulting from $q_y^2$ terms in Eq. (\ref{eqn:ek_fxy}).

\subsection{Thermal Hall conductivity}
In this subsection we discuss the temperature dependence of the thermal Hall conductivity. 
The formula in a spin-triplet superconductor with broken time-reversal symmetry is given by~\cite{sumiyoshi13} 
\begin{align}
\kappa_{xy}&=-\frac{1}{4\pi T}\int d\varepsilon \,\varepsilon^2 \Lambda ( \varepsilon) f'(\varepsilon),
\label{eqn:kappa}
\\
\Lambda ( \varepsilon)&=\frac{4\pi}{M}\sum_{\bm k n=\pm}{\rm Im}\left\{ \left\langle \frac{\partial u_{\bm kn}}{\partial k_x}\Bigg|\frac{\partial u_{\bm kn}}{\partial k_y}\right\rangle\right\}\theta(\varepsilon -E_{\bm kn}),\nonumber\\
\label{eqn:Lambda}
\end{align}
where $f'(\varepsilon )$ denotes the derivative of the Fermi-distribution function.  
The treatment of the numerical calculation is the same as that of the Chern number. 
In the low-temperature limit, the thermal Hall conductivity for the topological superconductors becomes linear in temperature as
\begin{align}
\kappa_{xy}\approx \kappa_{xy}^{L}=\frac{\pi N_C}{12}T. 
\end{align}

\begin{figure}[bt]
\begin{center}
\includegraphics[height=30mm]{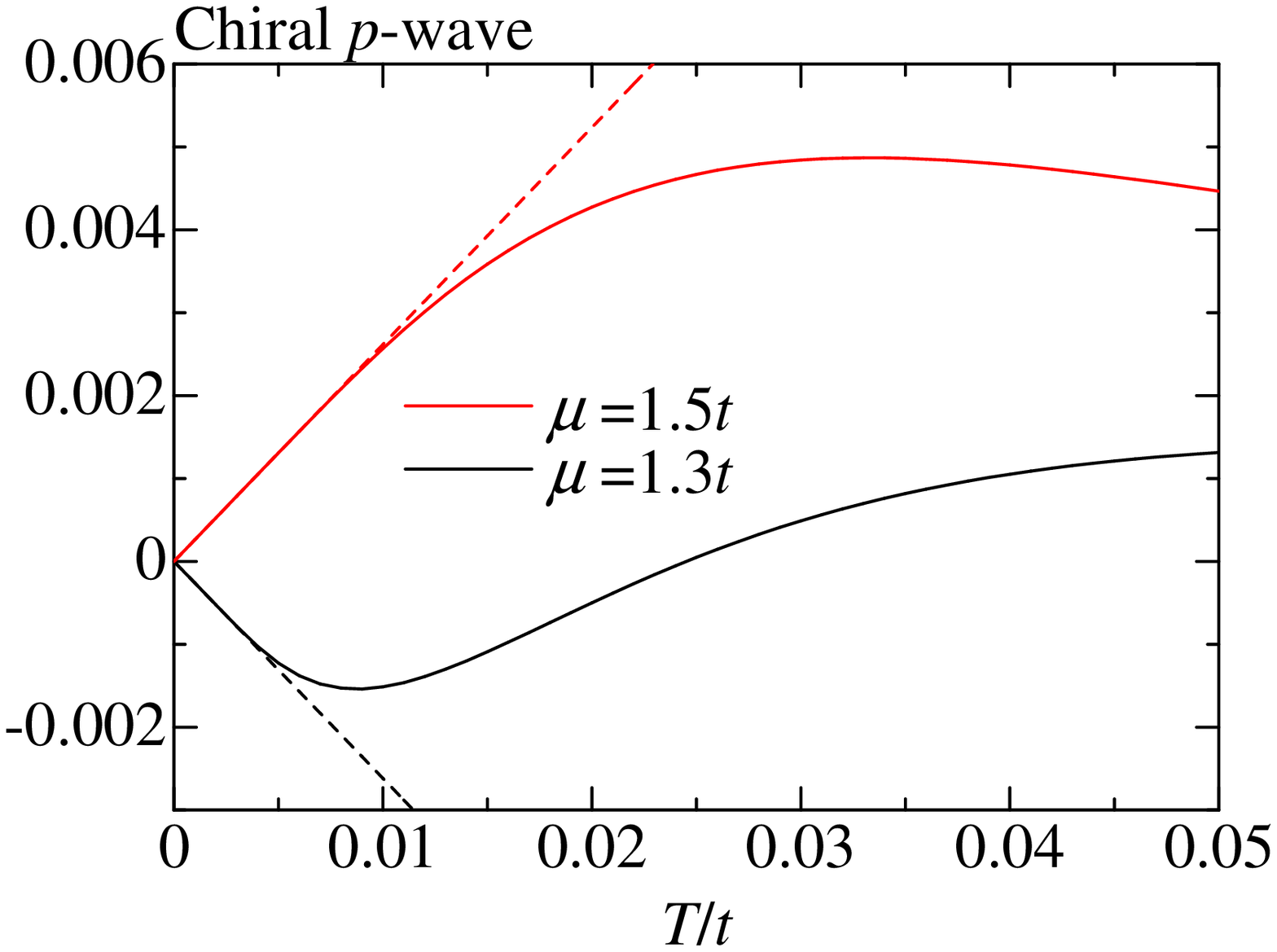}
\includegraphics[height=30mm]{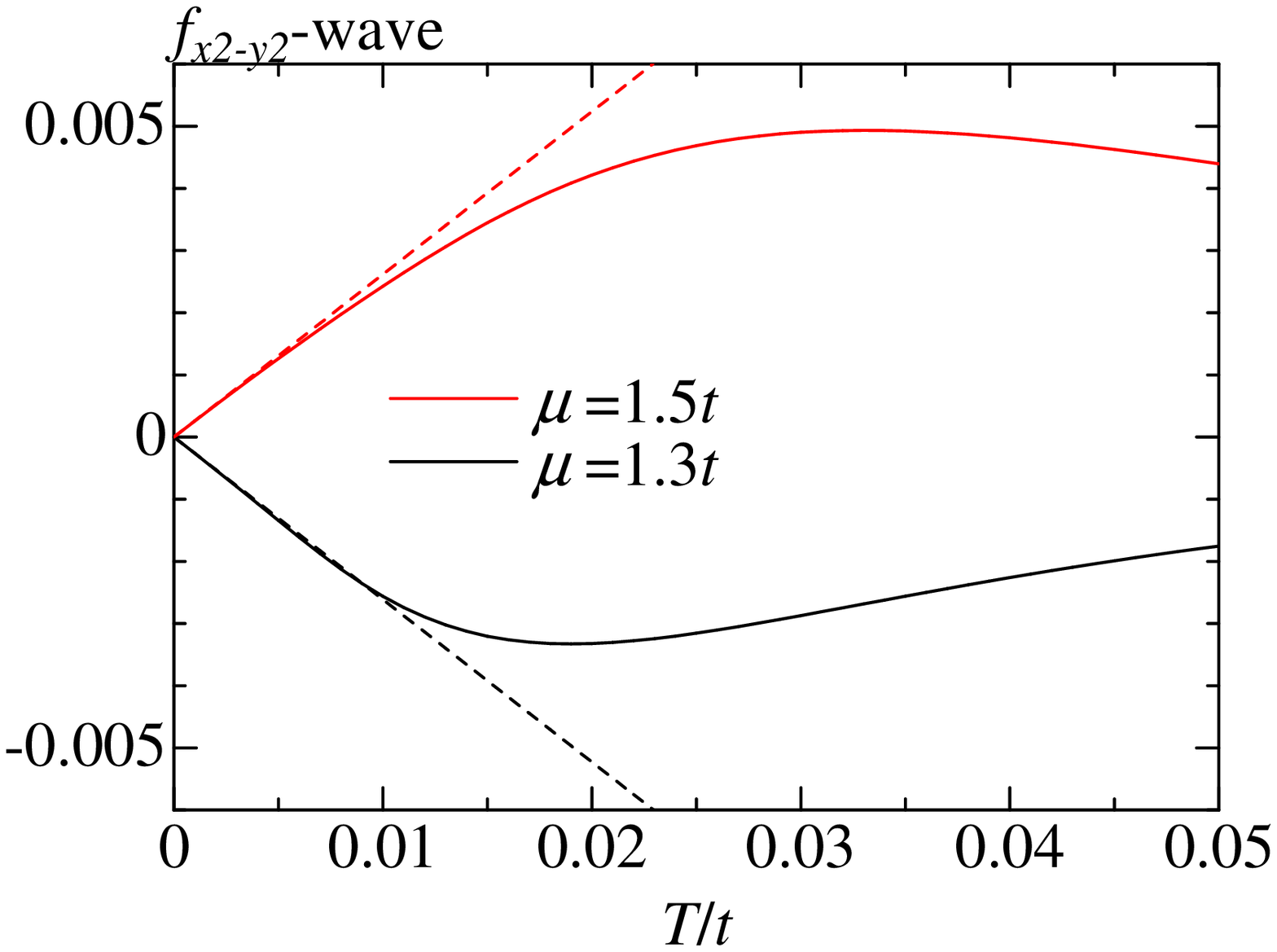}
\includegraphics[height=30mm]{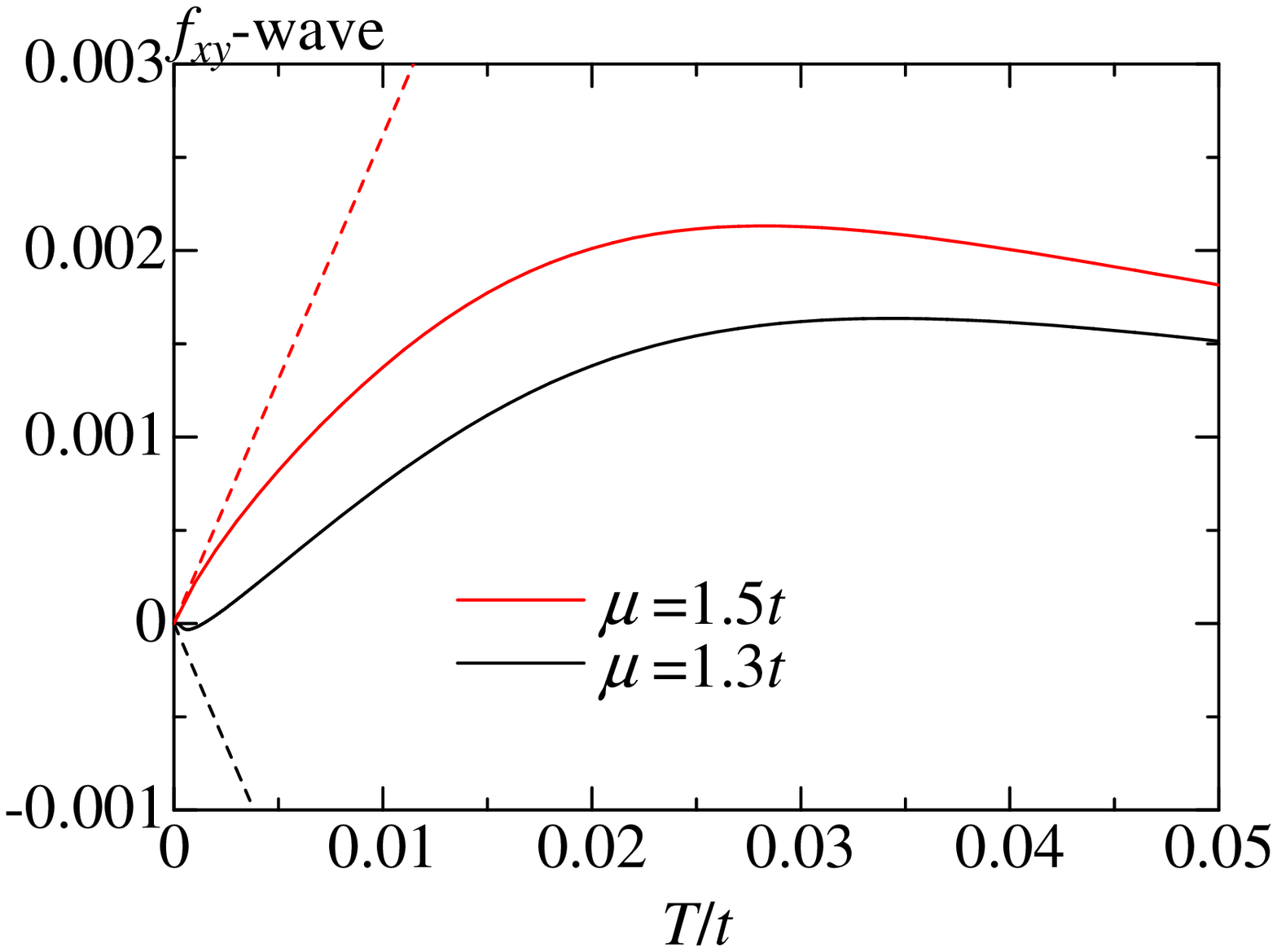}
\caption{(Color online) Temperature dependence of the thermal Hall conductivity for the chiral $p$-wave, the $f_{x^2-y^2}$-wave, and the $f_{xy}$-wave states, respectively.  The dashed lines represent $\kappa_{xy}^{L}$. }
\label{kxy}
\end{center}
\end{figure}

The temperature dependence of the thermal Hall conductivity $\kappa_{xy}$ is depicted in Fig. \ref{kxy}. 
In contrast to the chiral $p$-wave and the $f_{x^2-y^2}$-wave phases, $\kappa_{xy}$ in the $f_{xy}$-wave phase obviously deviates from $\kappa_{xy}^{L}$ even in the very low-temperature region, because $N_C$ is ill-defined. 
$\kappa_{xy}$ in the $f_{xy}$-wave phase does also not show the drastic change around the Lifshitz transition except in the extremely low-temperature region such that $\kappa_{xy}$ for the $f_{xy}$-wave state is less effective in detecting Fermi surface tuning related behavior, such as lattice distortion or strain effects, than for the other pairing states. 

\begin{figure}[b]
\begin{center}
\includegraphics[height=30mm]{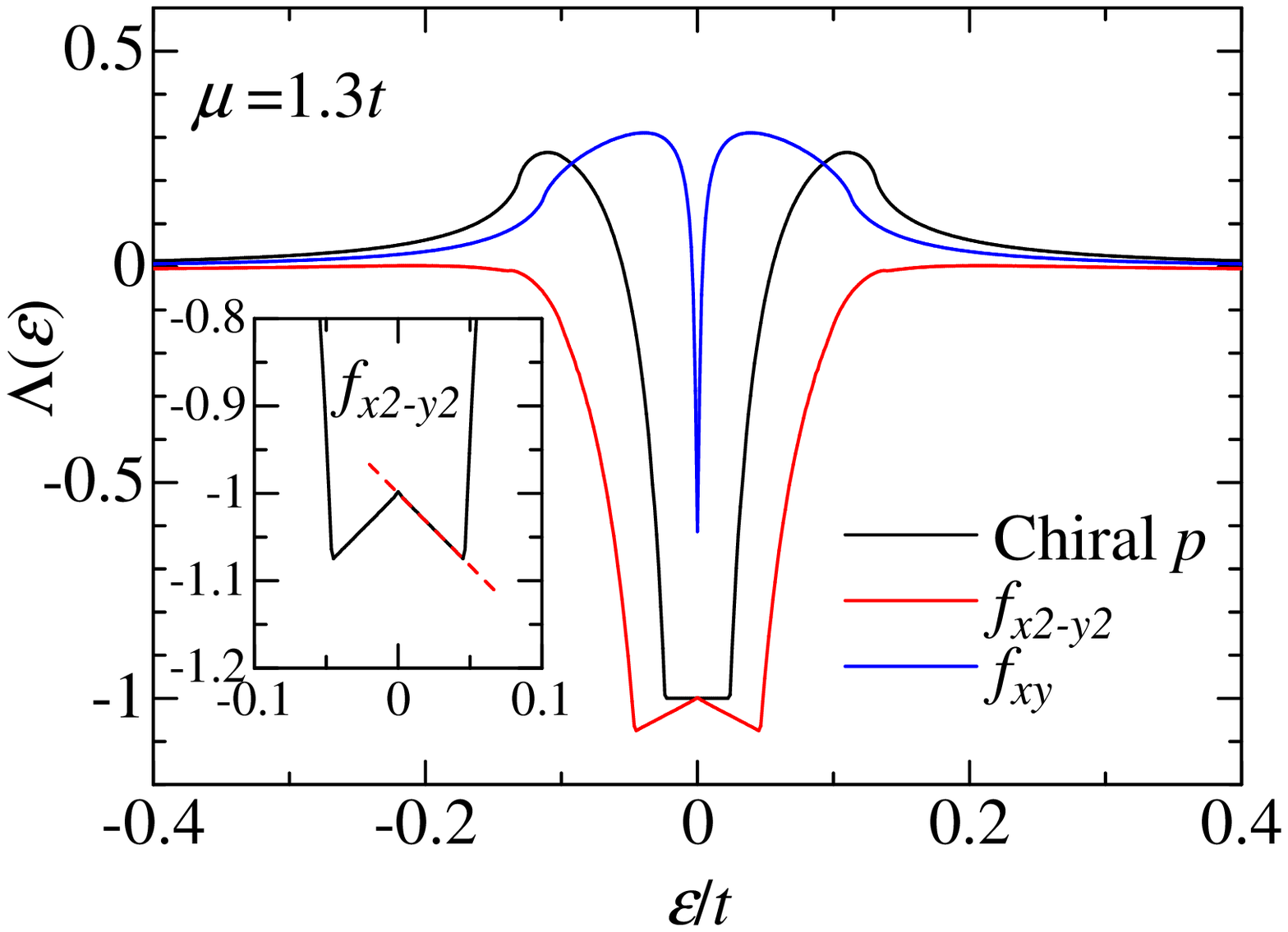}
\includegraphics[height=30mm]{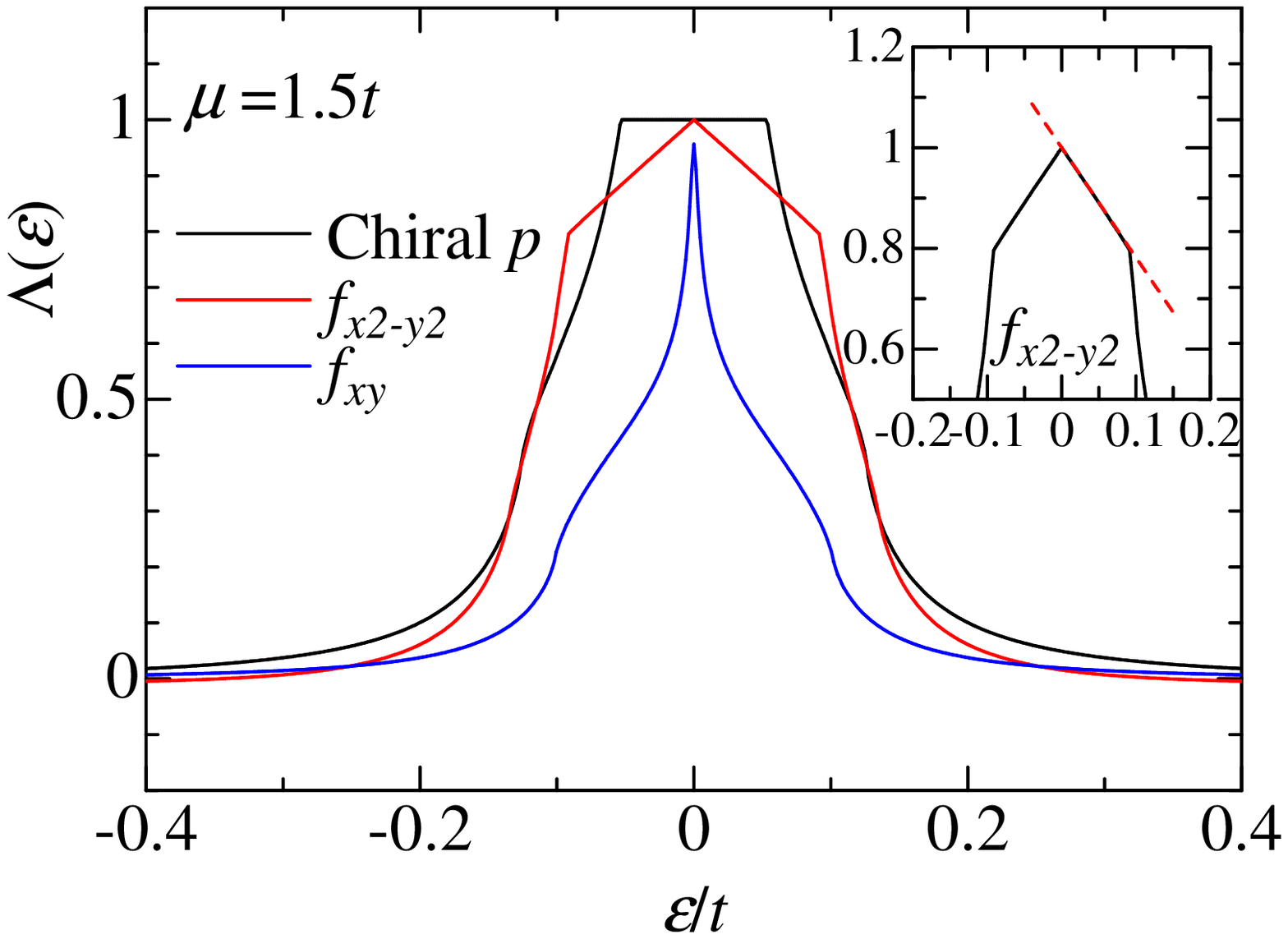}
\caption{(Color online) The structure of $\Lambda ( \varepsilon)$ with $\mu=1.3t$ (left panel) and $\mu=1.5t$ (right panel) for the chiral $p$-wave, the $f_{x^2-y^2}$-wave, and the $f_{xy}$-wave states, respectively.  
The insets show the magnification in the low-energy region for the $f_{x^2-y^2}$-wave phase. The red-dashed lines stand for linear fittings. 
}
\label{Lambda}
\end{center}
\end{figure}

Next, let us consider the structure of $\Lambda ( \varepsilon)$ which determines the temperature dependence of the thermal Hall conductivity, depicted in Fig. \ref{Lambda}. 
Note that $\Lambda ( \varepsilon)$ at the low-energy limit ($\varepsilon\rightarrow 0$) corresponds to $N_C$ in topological superconductors.  
The low-energy structures with the rectangular-like (the chiral $p$-wave and $f_{x^2-y^2}$-wave phases) and with the cusp ($f_{xy}$-wave phase) forms change drastically from convex downward to convex upward around $\mu=\mu_c$, whose structure dominates the low-temperature property of $\kappa_{xy}$. 

With increasing $|\varepsilon|$, $\Lambda ( \varepsilon)$ reveals a bending at the characteristic energy defined as $\varepsilon_0$, which corresponds to a minimum gap (pseudogap) near the van Hove points in the quasiparticle DOS for the chiral $p$-wave ($f_{x^2-y^2}$-wave) state~\cite{imai16}. 
For the larger $|\varepsilon|\gg \varepsilon_0$ region, $\Lambda ( \varepsilon)$ decreases towards zero, which depends on details of the model, and is almost independent of the topological aspect and the Lifshitz transition. 

In comparison to the chiral $p$-wave and the $f_{x^2-y^2}$-wave phases,  the central peak of $\Lambda ( \varepsilon)$ in the low-energy region for the $f_{xy}$-wave state is smaller and much narrower, so that the temperature dependence of the thermal Hall conductivity becomes less sensitive to the tuning of the chemical potential. 
Note that the constant $\xi $ is irrelevant to the temperature dependence of  $\kappa_{xy}$ because $\kappa_{xy}$ vanishes at $T\rightarrow 0$ and the contribution of the narrow central peak of $\Lambda ( \varepsilon)$ to $\kappa_{xy}$ is small at $T>0$. 

$\Lambda ( \varepsilon)$ for the $f_{x^2-y^2}$-wave state in the low-energy region has a linear slope, which can be represented analytically as
\begin{align}
\Lambda ( \varepsilon)&\approx N_C+\alpha \eta|\varepsilon |\>\>(|\varepsilon |\sim 0), \\ 
 \alpha &\equiv \frac{\cos k_0}{2 \sqrt{2}\Delta\sin^3 k_0}.  
\end{align}
This formula is obtained by means of the expansion in a power series with respect to wave vector around the gapless points [$\bm k=(k_0,k_0)$] in Eq. (\ref{eqn:Lambda}), whose fitting result is depicted in Fig. \ref{Lambda} insets. 
In the present parameter region, since $k_0$ is in the range of $\pi/2 < k_0<\pi$, the coefficient $\alpha$ changes continuously with increasing $\mu$ and is negative due to $\cos k_0<0$ and $\sin k_0>0$ regardless of the Lifshitz transition. 
Note that $N_C$ and $\alpha$ have same (opposite) signs for $\mu<\mu_c$ ($\mu>\mu_c$) and the sign of the Chern number also depends on the chirality $\eta$, so that the switching of $\eta$ generates the total sign change of $\Lambda ( \varepsilon)$. 

For an illustrative analysis of temperature dependence of the thermal Hall conductivity, we use a following function $\tilde{\Lambda} ( \varepsilon)$ for the chiral $p$-wave and the $f_{x^2-y^2}$-wave states instead of the exact $\Lambda ( \varepsilon)$
\begin{align}
\tilde{\Lambda} ( \varepsilon)&=\left\{
\begin{array}{cc}
N_C, &|\varepsilon |\leq \varepsilon_0\\
0, & \text{otherwise}
\end{array}
\right. \>(\text{chiral } p)
\label{eqn:lambda_p+ip}
\\
%
&=\left\{
\begin{array}{cc}
N_C+\alpha \eta|\varepsilon |, &|\varepsilon |\leq \varepsilon_0\\
0, & \text{otherwise}
\end{array}
\right. \>(f_{x^2-y^2}).
\label{eqn:lambda_fx2y2}
\end{align}

Substituting Eqs. (\ref{eqn:lambda_p+ip}) and (\ref{eqn:lambda_fx2y2}) into Eq. (\ref{eqn:kappa}), we obtain the following approximated thermal Hall conductivity $\tilde{\kappa}_{xy}$ analytically in the low-temperature region ($T\ll \varepsilon_0$) 
\begin{equation}
\tilde{\kappa}_{xy}=\left\{
\begin{array}{cc}
\kappa^L_{xy}-\gamma(T)e^{-\frac{\varepsilon_0}{T}}, & (\text{chiral } p)  \\
\kappa^L_{xy}+\kappa^Q_{xy}
-\left\{\gamma(T)+\gamma'(T)\right\}e^{-\frac{\varepsilon_0}{T}},
& (f_{x^2-y^2})
\end{array}
\label{eqn:approx_kappa}
\right.
\end{equation}
where
\begin{align}
\kappa^Q_{xy}&=\frac{9\zeta(3)}{4\pi}\alpha \eta T^2,\\
\gamma(T)&=\frac{TN_C}{\pi}\left\{ 2\left( \frac{\varepsilon_0}{2T}\right)^2+2\left( \frac{\varepsilon_0}{2T}\right)+1 \right\},\\
\gamma'(T)&=\frac{2T^2\alpha \eta}{\pi}\left\{ 2\left( \frac{\varepsilon_0}{2T}\right)^3+3\left( \frac{\varepsilon_0}{2T}\right)^2+3\left( \frac{\varepsilon_0}{2T}\right)+\frac{3}{2}\right\}. \nonumber \\
\label{eqn:gamma}
\end{align}
$\zeta(s)$ is the Riemann $\zeta$ function with $\zeta(3)\approx 1.202$. 

In this analysis, the thermal Hall conductivity is expected to be composed of the $T$-linear term and an exponentially small correction in the chiral $p$-wave phase. 
In addition to the contribution from these terms, a quadratic term in temperature $\kappa^Q_{xy}$ appears in the $f_{x^2-y^2}$-wave phase. 

\begin{figure}[t]
\begin{center}
\includegraphics[height=30mm]{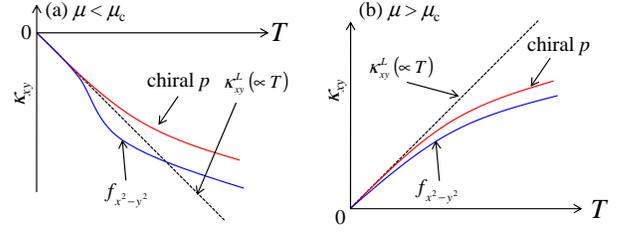}
\caption{(Color online) Schematic picture of the temperature dependence of the thermal Hall conductivity
for $\mu<\mu_c$ (left panel) and $\mu>\mu_c$ (right panel).
}
\label{kappa_sp}
\end{center}
\end{figure}

While the thermal Hall conductivity deviates from the $T$-linear term monotonically with increasing temperature in the chiral $p$-wave phase, in the $f_{x^2-y^2}$-wave phase we find a non-monotonic behavior due to the $T^2$ term ($\kappa^Q_{xy}$) resulting from the linear dispersions around the gapless points for $\mu<\mu_c$. The schematic behavior is depicted in Fig. \ref{kappa_sp}. 

In the case of $\mu>\mu_c$, since the sign of $N_C$ is different from that of $\alpha$, the thermal Hall conductivity shows similar temperature dependence in both the chiral $p$-wave and the $f_{x^2-y^2}$-wave phases. In comparison with the chiral $p$-wave phase, however, a rapid deviation from the $T$-linear term occurs for the $f_{x^2-y^2}$-wave phase due to the contribution from the $T^2$ term.  
Note that we consider here the chirality $\eta=+1$ case, and the thermal Hall conductivity with $\eta=-1$ leads merely to the sign inversion of that with $\eta=+1$.

\begin{figure}[tb]
\begin{center}
\includegraphics[height=30mm]{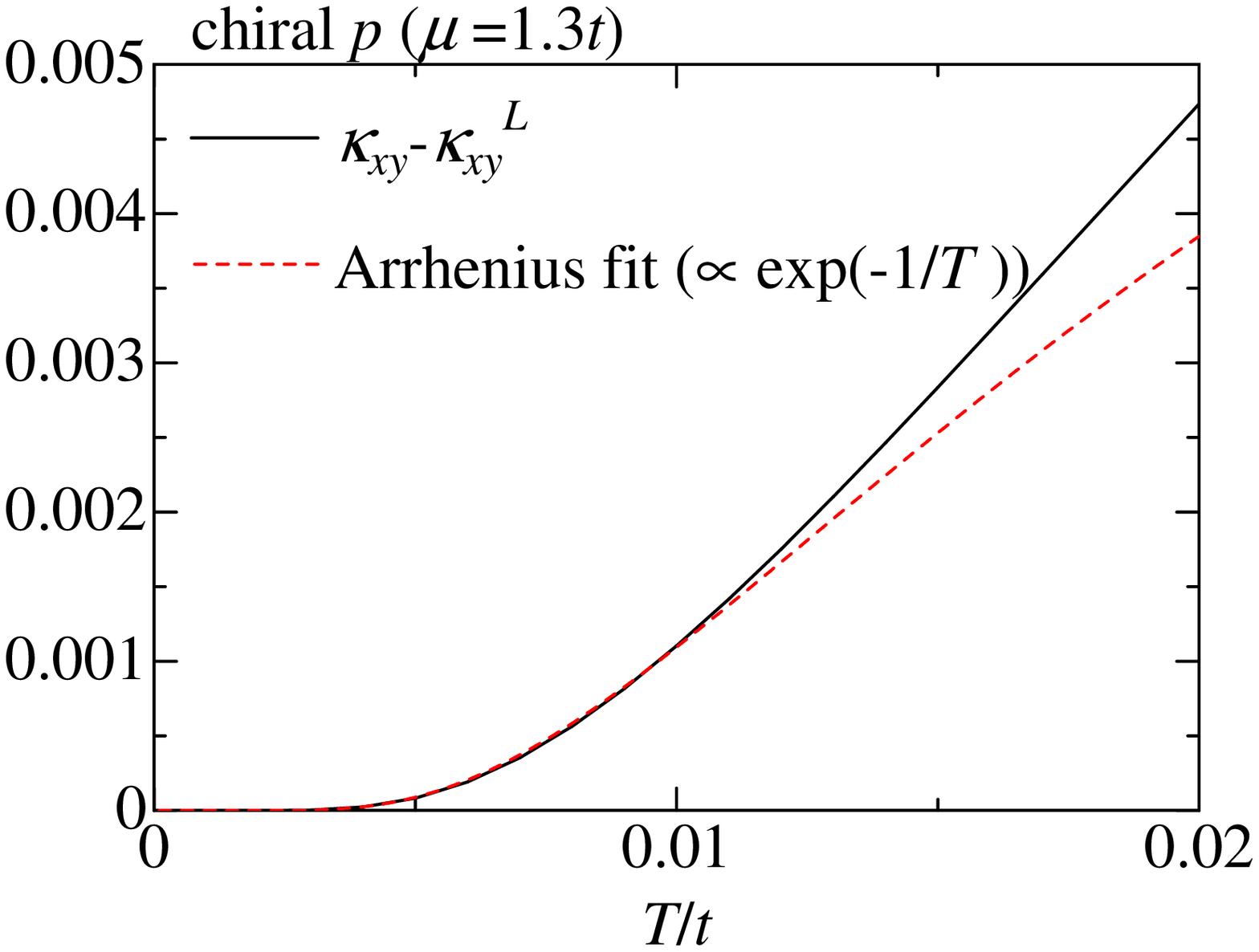}
\includegraphics[height=30mm]{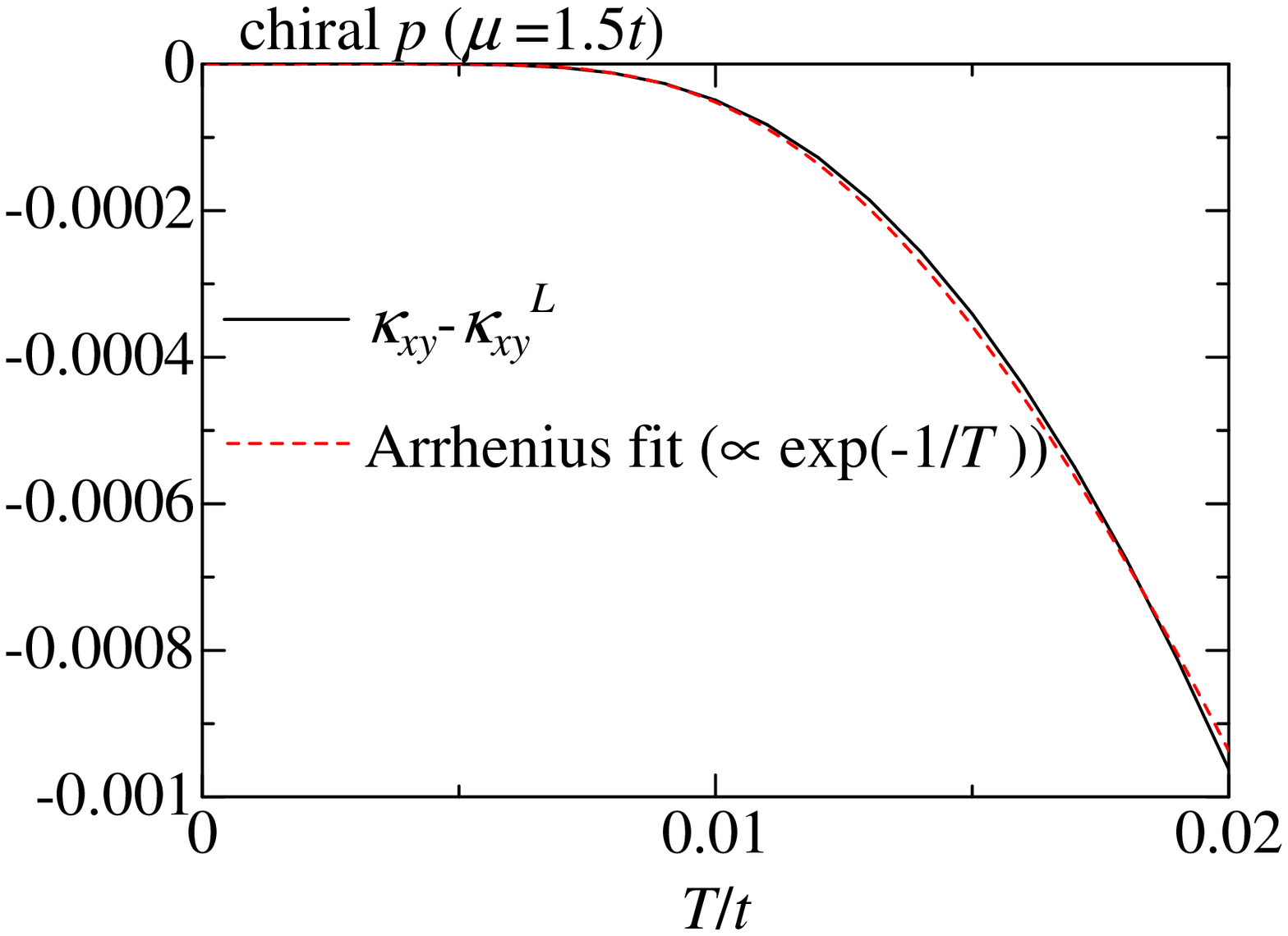}
\includegraphics[height=30mm]{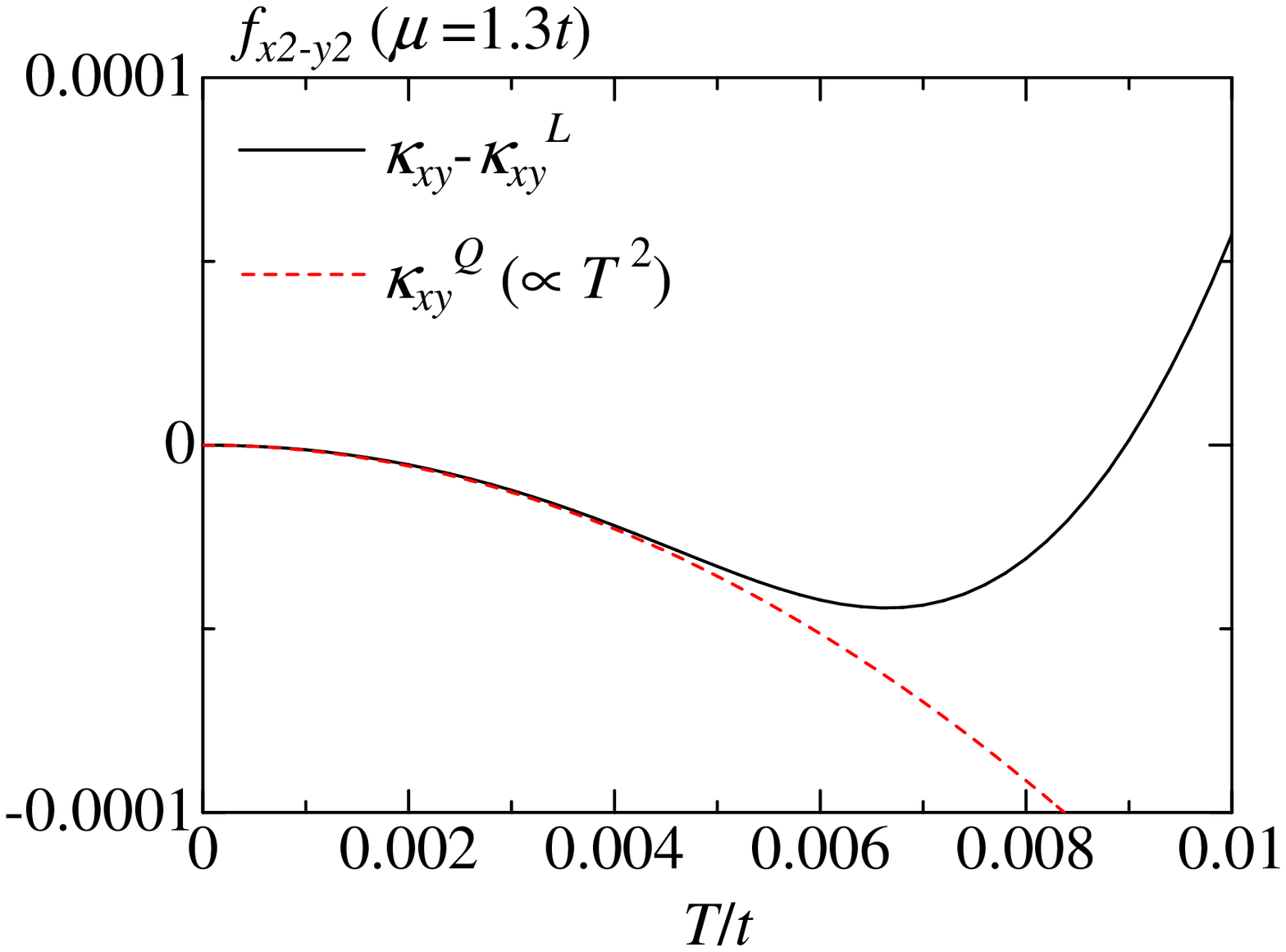}
\includegraphics[height=30mm]{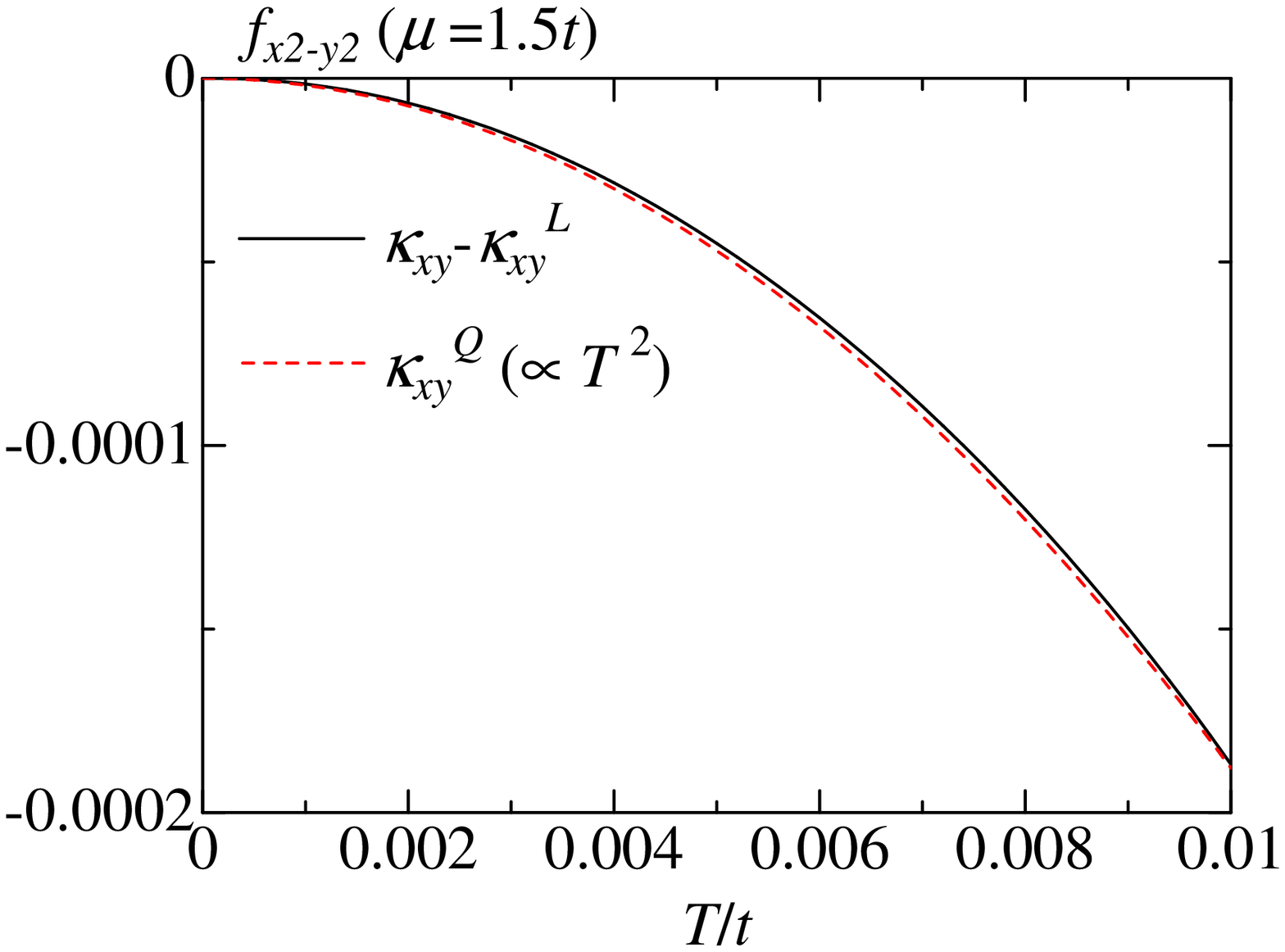}
\caption{(Color online) The difference between $\kappa_{xy}$ and $\kappa^L_{xy}$ as a function of temperature for $\mu<\mu_c$ (left panel) and $\mu>\mu_c$ (right panel). 
Upper (lower) panels represent the chiral $p$-wave ($f_{x^2-y^2}$-wave) phase. 
The dashed lines stand for the exponential fitting estimated by the Arrhenius fit (upper panels) and $T^2$ term [$\kappa^Q_{xy}= (9\zeta(3)/4\pi)\alpha T^2$] in Eq. (\ref{eqn:approx_kappa}) (lower panels), respectively. 
}
\label{fit}
\end{center}
\end{figure}

Thus, deviations from the $T$-linear term are expected to show the exponential and the $T^2$ behaviors in the low-temperature region for the chiral $p$-wave and the $f_{x^2-y^2}$-wave phases, respectively. 
Figure \ref{fit} displays $\kappa_{xy}-\kappa^L_{xy}$ for the chiral $p$-wave and the $f_{x^2-y^2}$-wave states where $\kappa_{xy}$ is obtained numerically. 
The dashed-lines in the the chiral $p$-wave phase are obtained by an Arrhenius fit as $\kappa_{xy}-\kappa^L_{xy}\approx \beta e^{-\frac{\delta}{T}}$, 
where $\beta$ and $\delta$ are the fitting parameters, and the latter corresponds to the minimum superconducting gap $\varepsilon_0$~\cite{imai16}. 
A power analysis concerning $T$ for the $f_{x^2-y^2}$-wave phase essentially corresponds to $\kappa^Q_{xy}=(9\zeta(3)/4\pi)\alpha T^2$ in Eq. (\ref{eqn:approx_kappa}). 
These results indicate that the temperature dependence of the thermal Hall conductivity is well described with the present approximation. 

With further increasing temperature in the $f_{x^2-y^2}$-wave phase, the exponential term becomes dominant in the thermal Hall conductivity, which is well described as 
$\kappa_{xy}-\kappa^L_{xy}-\kappa^Q_{xy}\approx \beta' e^{-\frac{\delta'}{T}}$. 
Thus $\delta'$ obtained by the Arrhenius fit reproduces the pseudogap amplitude $\varepsilon_0$. 

We stress that in particular, the chemical potential of the bulk Sr$_2$RuO$_4$ is expected to be slightly smaller than $\mu_c$, and the deviation from the $T$-linear term ($\kappa_{xy}-\kappa^L_{xy}$) for $\mu<\mu_c$ shows a clear difference between the chiral $p$-wave and the $f_{x^2-y^2}$-wave phases. This result may enable us to distinguish the pairing symmetries in Sr$_2$RuO$_4$.  

\section{Summary}
We have investigated the temperature dependence of the thermal Hall conductivity in the spin-triplet superconducting states with broken time-reversal symmetry. 
By means of the two-dimensional tight-binding model describing the $\gamma$ band, we calculate the thermal Hall conductivity for the chiral $p$-wave state and the $f$-wave ($f_{x^2-y^2}$ and $f_{xy}$) states. Since the former has a nodeless superconducting gap, the topological properties emerge clearly even in the finite temperature region, and the Chern number changes concomitantly with the Lifshitz transition. 
The two $f$-wave phases have nodes in the gap functions yielding gapless excitations. 
While the $f_{x^2-y^2}$-wave state possesses linear dispersions around the gapless points in the bulk system, the Chern number becomes integer only at absolute zero temperature. 
On the other hand,  in the $f_{xy}$-wave phase a  finite DOS at $\epsilon \sim 0$ exists, originating from the quadratic dispersion with respect to wave vector around the gapless points. Here the Chern number deviates from integer even at absolute zero temperature.  

In the topological superconducting phase, it is well known that a $T$-linear term proportional to the Chern number appears in the thermal Hall conductivity. 
In the chiral $p$-wave and the $f_{x^2-y^2}$-wave phases the thermal Hall conductivity changes drastically due to the sign switching of the Chern number around the Lifshitz transition. 
Compared with these pairing phases, the thermal Hall conductivity in the $f_{xy}$-wave phase is essentially insensitive to the topological properties and the change of the Fermi surface topology.  

While the thermal Hall conductivity in the chiral $p$-wave phase consists of the temperature linear and the exponential correction term, the $f_{x^2-y^2}$-wave phase lead to a quadratic correction term with respect to temperature due to the linear dispersions. 
The coefficient of the quadratic term is not connected with the topological character and the Lifshitz transition, but the band structure around the gapless points. 
Thus, since the sign of the coefficient corresponds to that of the Chern number in the parameter region of Sr$_2$RuO$_4$, the thermal Hall conductivity shows obviously different behavior for the chiral $p$-wave state and the $f_{x^2-y^2}$-wave states in the low-temperature region. 
It is important to notice that although both $f$-wave states are "nodal" superconductors, the position of the nodes within the Brillouin zone has a strong influence on the low-energy quasiparticle spectrum and the low-temperature behavior, not only for the thermal Hall effect, but also for other quantities such as the specific heat or the London penetration depth. 
This result may help to distinguish different chiral pairing states. 

The contributions from other types of bands, such as the $\alpha$ and $\beta$ bands, whose Fermi surfaces are more removed from the van Hove points, are not taken into account in the present study. 
The $f_{x^2-y^2}$-wave state would give rise to the gapless excitations on those bands, which may generate a power-law behavior in the temperature dependencs for the thermal Hall conductivity in contrast to the chiral $p$-wave phase.  This multi-band effect is an interesting open problem to be addressed in the future.

\begin{acknowledgments}
We are grateful to A. Bouhon, J. Goryo, Y. Maeno, T. Neupert, T. Saso, A. Schnyder, Y. Yanase and N. Yoshioka for many helpful discussions. 
This work is supported by the Swiss National Science Foundation and JSPS KAKENHI Grant Number JP15K13507. 
YI is grateful for hospitality by the Pauli Center for Theoretical Studies of ETH Z\"urich. 
\end{acknowledgments}

\appendix*
\section{}
In this appendix, the derivation of the approximate thermal Hall conductivity in Eq. (\ref{eqn:approx_kappa}) is given as follows, 
\begin{align}
\tilde{\kappa}_{xy}&=-\frac{1}{4\pi T}\int d\varepsilon \,\varepsilon^2 \tilde{\Lambda} ( \varepsilon) f'(\varepsilon)\nonumber\\
%
%
%
&=\frac{1}{4\pi T^2}\int_{-\varepsilon _0}^{\varepsilon _0} d\varepsilon \,
\frac{N_C \varepsilon^2+\alpha \varepsilon^2|\varepsilon |}{4\cosh^2\left(\frac{\beta \varepsilon}{2}\right)} 
\nonumber\\
%
%
%
&=\frac{N_CT}{\pi}  \int_{0}^{x_0} dx \,\frac{x^2}{\cosh^2x}
+\frac{2\alpha T^2}{\pi} \int_{0}^{x_0} dx \,\frac{x^3}{\cosh^2x},\nonumber\\ 
\end{align}
where $x_0\equiv \varepsilon_0/(2T)$. 
%
The integrations are carried out as follows, 
\begin{align}
\int_{0}^{x_0} dx \,\frac{x^2}{\cosh^2x}&=\int_{0}^{\infty} dx \,\frac{x^2}{\cosh^2x}-\int_{x_0}^{\infty} dx \,\frac{x^2}{\cosh^2x}
\nonumber \\
%
%
&\approx \frac{\pi^2}{12}-\int_{x_0}^{\infty} dx \,4x^2 e^{-2x}\>\>(x_0\gg 1)\nonumber\\
%
%
&=\frac{\pi^2}{12}-e^{-2x_0}\left( 4x_0^2+4x_0+2 \right), \\
\int_{0}^{x_0} dx \,\frac{x^3}{\cosh^2x}&=\int_{0}^{\infty} dx \,\frac{x^3}{\cosh^2x}-\int_{x_0}^{\infty} dx \,\frac{x^3}{\cosh^2x}
\nonumber \\
%
%
&\approx \frac{9}{8}\zeta(3)-\int_{x_0}^{\infty} dx \,4x^3 e^{-2x}\>\>(x_0\gg 1)\nonumber\\
%
%
&=\frac{9}{8}\zeta(3)-e^{-2x_0}\left( 2x_0^3+3x_0^2+3x_0+\frac{3}{2} \right). \nonumber\\ 
\end{align}
%
Finally, we obtain Eqs. (\ref{eqn:approx_kappa})-(\ref{eqn:gamma}). 

\bibliography{paper.bib}
\end{document}